\documentclass[a4paper,12pt]{article}
\pdfoutput=1
\usepackage[bookmarks=false]{hyperref}
\usepackage{amssymb,amsmath}
\usepackage{graphicx,subfigure}

\newlength{\dinwidth}
\newlength{\dinmargin}
\begin{document}

\title{\bf $B_s^0-\bar B_s^0$ mixing in a family non-universal $Z^{\prime}$ model revisited}
\bigskip

\author{Xin-Qiang Li$^{1,2}$, Yan-Min Li$^{1}$, Gong-Ru Lu$^{1}$ and Fang Su$^{3}$\\
{ $^1$\small Department of Physics, Henan Normal University, Xinxiang, Henan 453007, P.~R. China}\\
{ $^2$\small IFIC, Universitat de Val\`encia-CSIC, Apt. Correus 22085, E-46071 Val\`encia, Spain}\\
{ $^3$\small Institute of Particle Physics, Huazhong Normal University, Wuhan, Hubei 430079, P.~R. China}}

\date{}
\maketitle
\bigskip \bigskip
\maketitle
\vspace{-1.5cm}

\begin{abstract}
{\noindent}Motivated by the very recent measurements performed at the LHCb and the Tevatron of the $B_s^0-\bar B_s^0$ mixing, in this paper we revisit it in a family non-universal $Z^{\prime}$ model, to check if a simultaneous explanation for all the mixing observables, especially for the like-sign dimuon charge asymmetry observed by the D0 collaboration, could be made in such a specific model. In the first scenario where the $Z^\prime$ boson contributes only to the off-diagonal element $M_{12}^s$, it is found that, once the combined constraints from $\Delta M_s$, $\phi_s$ and $\Delta \Gamma_s$ are imposed, the model could not explain the measured flavour-specific CP asymmetry $a_{fs}^s$, at least within its $1\sigma$ ranges. In the second scenario where the NP contributes also to the absorptive part $\Gamma_{12}^s$ via tree-level $Z^\prime$-induced $b\to c\bar{c}s$ operators, we find that, with the constraints from $\Delta M_s$, $\phi_s$ and the indirect CP asymmetry in $\bar{B}_d\to J/\psi K_S$ taken into account, the present measured $1\sigma$ experimental ranges for $a_{fs}^s$ could not be reproduced too. Thus, such a specific $Z^\prime$ model with our specific assumptions could not simultaneously reconcile all the present data on $B_s^0-\bar B_s^0$ mixing. Future improved measurements from the LHCb and the proposed superB experiments, especially of the flavour-specific CP asymmetries, are expected to shed light on the issue.

\end{abstract}

\newpage

\section{Introduction}
\label{Sec:intro}

Within the Standard Model~(SM), the flavour-changing neutral current~(FCNC) processes occur only at the loop level, and are therefore a very sensitive probe of new physics~(NP) beyond the SM. In this direction, an outstanding role is played by the $B_s^0-\bar B_s^0$ mixing, the phenomenon of which is described by two off-diagonal elements, $M_{12}^s$ of the mass and $\Gamma_{12}^s$ of the decay matrix. These two complex parameters can be fully determined by the following four observables: the mass difference $\Delta M_s$, the decay width difference $\Delta\Gamma_s$, the CP-violating phase $\phi_s$, as well as the flavour-specific CP asymmetry $a_{fs}^s$ in $B_s$-meson decays~\cite{Ball:2006xx,Asner:2010qj}.

The mass difference $\Delta M_s$ was first measured at the Tevatron~\cite{Abazov:2006dm,Abulencia:2006mq}, with the most precise published value given by the CDF collaboration~\cite{Abulencia:2006mq}
\begin{equation}\label{deltams-cdf}
\Delta M_s^{\rm CDF}=(17.77\pm 0.10({\rm stat.}) \pm 0.07({\rm syst.}))~{\rm ps}^{-1}\,.
\end{equation}
Employing a data sample of $340{\rm pb}^{-1}$ taken in 2011, the LHCb collaboration has recently updated its analysis made in 2010~\cite{Aaij:2011qx}, and the most precise value of $\Delta M_s$ is found to be~\cite{LHCbnote50}
\begin{equation}\label{deltams-lhcb}
\Delta M_s^{\rm LHCb}=(17.725\pm 0.041({\rm stat.}) \pm 0.026({\rm syst.}))~{\rm ps}^{-1}\,.
\end{equation}
Both of these measurements agree quite well with the corresponding SM prediction~\cite{Lenz:2011zz,Lenz:2011ti,Lenz:2010gu}
\begin{equation}\label{deltams-sm}
\Delta M_s^{\rm SM}=(17.3\pm 2.6)~{\rm ps}^{-1}\,.
\end{equation}

The decay width difference $\Delta\Gamma_s$ and the CP-violating phase $\phi_s$ can be simultaneously determined from an angular analysis of the decay $B_s\to J/\psi \phi$. Both the CDF~\cite{CDF:2011af} and D0~\cite{Abazov:2011ry} collaborations have recently updated their previous analysis based on $1.35{\rm fb}^{-1}$~\cite{Aaltonen:2007he} and $2.8{\rm fb}^{-1}$~\cite{:2008fj} of data sample, respectively. While the two measurements agree within uncertainties, the CDF result constrains $\phi_s$ to a narrower region, and we therefore quote~\cite{CDF:2011af}
\begin{align} \label{betas-deltagammas-tevatron}
\phi_s^{\rm CDF} &\in [-1.04,-0.04] \cup [-3.10,-2.16]~{\rm rad}\,, \nonumber \\[0.2cm]
\Delta\Gamma_s^{\rm CDF} &= (0.075 \pm 0.035({\rm stat.}) \pm 0.006({\rm syst.}))~{\rm ps}^{-1}\,,
\end{align}
given at $68\%$ confidence level~(C.L.). It is noted that these measurements have been superseded by the recent LHCb analysis~\cite{LHCb:2011aa}. Combining the two channels $B_s\to J/\psi \phi$~\cite{LHCb:2011aa} and $B_s\to J/\psi f_0$~\cite{LHCb:2011ab}, the LHCb collaboration presents the most precise measurement at $68\%$ C.L.~\cite{LHCb:2011ab}
\begin{align} \label{betas-deltagammas-lhcb}
\phi_s^{\rm LHCb} &=(0.07 \pm 0.17({\rm stat.}) \pm 0.06({\rm syst.}))~{\rm rad}\,, \nonumber \\[0.2cm]
\Delta\Gamma_s^{\rm LHCb} &= (0.123 \pm 0.029({\rm stat.}) \pm 0.011({\rm syst.}))~{\rm ps}^{-1}\,.
\end{align}
Compared to the corresponding SM predictions~\cite{Lenz:2011zz,Lenz:2011ti,Lenz:2010gu}
\begin{align} \label{betas-deltagammas-sm}
\phi_s^{\rm SM} &= -2\arg[-(V_{ts}V_{tb}^\ast)/(V_{cs}V_{cb}^\ast)] = (-0.037\pm 0.002)~{\rm rad}\,, \nonumber \\[0.2cm]
\Delta\Gamma_s^{\rm SM} &= (0.087 \pm 0.021)~{\rm ps}^{-1}\,,
\end{align}
these improved measurements show a better agreement with the SM expectations; especially, no evidence for a NP phase in $B_s^0-\bar B_s^0$ mixing is found by the LHCb collaboration~\cite{LHCb:2011aa,LHCb:2011ab}.

The flavour-specific CP asymmetry $a_{fs}^s$ can be extracted from a measurement of the like-sign dimuon charge asymmetry $A_{\rm sl}^b$ of semileptonic b-hadron decays~\cite{Grossman:2006ce}. Employing a data sample of $9.0{\rm fb}^{-1}$ of $p\bar p$ collisions, the D0 collaboration has recently updated its previous analysis~\cite{Abazov:2010hv}, and the new measurement reads~\cite{Abazov:2011yk}
\begin{align} \label{Aslb-D0}
A_{\rm sl}^{b,\rm D0} &= (-0.787 \pm 0.172({\rm stat.}) \pm 0.093({\rm syst.}))\%\,, \nonumber\\[0.2cm]
a_{fs}^{s,\rm D0} &=(-1.81\pm 1.06)\%\,,
\end{align}
which, compared to the corresponding SM predictions~\cite{Lenz:2011zz,Lenz:2011ti,Lenz:2010gu}
\begin{align}
A_{\rm sl}^{b,\rm SM} &=(-2.0\pm 0.3)\times 10^{-4}\,, \nonumber\\[0.2cm]
a_{fs}^{s,\rm SM} &=(1.9\pm 0.3)\times 10^{-5}\,.
\end{align}
differ by about $3.9\sigma$ and $1.7\sigma$, respectively.

Motivated by the above observations, it is interesting to investigate if a specific NP model, while satisfy the constraints from $\Delta M_s$ and $\phi_s$, could also simultaneously explain the measured values of $\Delta \Gamma_s$ and $a_{fs}^{s}$. It is also expected that the combined constraints from these observables could provide further information about the model parameter space.

As is well-known, a $Z^{\prime}$ boson with family non-universal, flavour-changing couplings could arise in many well-motivated extensions of the SM with an additional ${\rm U(1)}^{\prime}$ gauge symmetry~\cite{Langacker:2008yv,Langacker:2000ju,Barger:2009eq}. Searching for such an extra $Z^{\prime}$ boson is an important mission of the Tevatron~\cite{Carena:2004xs} and LHC~\cite{Rizzo:2006nw} experiments. Performing constraints on the new $Z^{\prime}$ couplings through low-energy precise processes is, on the other hand, very crucial and complementary for these direct searches. It is interesting to note that such a family non-universal $Z^{\prime}$ model could bring new CP-violating phases beyond the SM and have a large effect on many FCNC processes~\cite{Langacker:2008yv,Langacker:2000ju,Barger:2009eq}, such as the $B_s-\bar{B}_s$ mixing~\cite{Chang:2009tx,BsBsbarmixing-old,BsBsbarmixing-new,Bobeth:2011st,Alok:2010ij}, as well as some rare~\cite{rareB} and hadronic B-meson decays~\cite{hadronicB}.

Thus, in this paper we shall revisit the $B_s-\bar{B}_s$ mixing in a family non-universal $Z^{\prime}$ model, and check if it could simultaneously explain the measured values of $\Delta M_s$, $\Delta \Gamma_s$, $\phi_s$, as well as $a_{fs}^{s}$, updating our previous analysis made in Ref.~\cite{Chang:2009tx}, where only effects of the flavour-changing left-handed $Z^\prime$ couplings, with all the right-handed couplings being flavour-diagonal, on $\Delta M_s$ and $\phi_s$ were considered. Furthermore, motivated by the like-sign dimuon charge asymmetry observed by the D0 collaboration~\cite{Abazov:2011yk}, we shall also consider a scenario with sizable $Z^\prime$ contribution to the off-diagonal element $\Gamma_{12}^s$ coming from the four-quark operators of the form $(\bar s b)_{V-A}(\bar c c)_{V \pm A}$.

Our paper is organized as follows. In Sec.~\ref{Sec:framework}, we first recapitulate the theoretical framework for $B_s^0-\bar B_s^0$ mixing, and then discuss the $Z^{\prime}$-boson contribution. In Sec.~\ref{Sec:numericalresult}, we give our detailed numerical results and discussions. Our conclusions are made in Sec.~\ref{Sec:conclusion}. The relevant input parameters are collected in the appendix.

\section{Theoretical framework for $B_s^0-\bar B_s^0$ mixing}
\label{Sec:framework}

In this section, we shall first recapitulate the theoretical framework for $B_s^0-\bar B_s^0$ mixing within the SM. The family non-universal $Z^{\prime}$ model, as well as its contribution to the off-diagonal elements $M_{12}^s$ and $\Gamma_{12}^s$, which govern all the observables in $B_s^0-\bar B_s^0$ mixing, are then discussed.

\subsection{Observables in $B_s^0-\bar B_s^0$ mixing}
\label{Sec:general}

The phenomenon of $B_s^0-\bar B_s^0$ mixing is described by a Schr\"{o}dinger equation~\cite{Ball:2006xx,Lenz:2010gu}
\begin{equation}
i \frac{d}{dt}\binom{|B_s(t)\rangle}{|\bar B_s(t)\rangle} = (M^s-\frac{i}{2} \Gamma^s)\binom{|B_s(t)\rangle}{|\bar B_s(t)\rangle}\,,
\end{equation}
with the mass matrix $M^s=M^{s\dagger}$ and the decay matrix $\Gamma^s=\Gamma^{s\dagger}$. By diagonalizing $M^s-\frac{i}{2} \Gamma^s$, one can obtain the two mass eigenstates $|B_{L}\rangle$, $|B_{H}\rangle$ with masses $M_{L}$, $M_{H}$ and decay rates $\Gamma_{L}$, $\Gamma_{H}$. Here $L$ and $H$ indicate the light and the heavy state, respectively. The mass and the width difference between the two states $|B_{H}\rangle$ and $|B_{L}\rangle$ are then defined, respectively, as~\cite{Ball:2006xx,Lenz:2010gu}
\begin{align}\label{eq:deltams-deltagammas-def}
\Delta M_s &\equiv M_{H}-M_{L}=2 |M_{12}^s|\,, \nonumber \\[0.2cm]
\Delta \Gamma_s &\equiv \Gamma_{L}-\Gamma_{H}=2 |\Gamma_{12}^s|\cos\Phi_s\,,
\end{align}
where $\Phi_s \equiv \arg(-M_{12}^s/\Gamma_{12}^s)$ is the CP-violating phase, and numerically irrelevant corrections that are proportional to $(\Gamma_{12}^s/M_{12}^s)^2$ have been neglected here.

The CP-violating phase $\phi_s$ appears in $b\to c\bar{c}s$ decays of $B_s$ meson, taking possible mixing effect into account. Taking the decay $B_s\to J/\psi\phi$ as an example, since it is dominated by the tree-level $b\to c\bar{c}s$ transition that is real within the SM, the measured CP-violating phase $\phi_s$ is a direct probe of the phase of $B_s^0-\bar B_s^0$ mixing. Assuming that there are no NP contributions to the decay amplitude\footnote{NP effects that contribute to the off-diagonal element $\Gamma_{12}^s$ might also give a contribution to the decay~\cite{Bobeth:2011st,Alok:2010ij,Chiang:2009ev}. For convenience, in this paper we shall not consider this case for $B_s$ meson.}, we can therefore write~\cite{Lenz:2010gu}
\begin{align}\label{eq:phis-def}
\phi_s \equiv \arg M_{12}^s=\phi_s^{\rm SM}+\phi_s^{\rm NP}\,,
\end{align}
with the second term denoting the NP phase contributing to the off-diagonal element $M_{12}^s$.

In terms of the two off-diagonal elements $M_{12}^s$ and $\Gamma_{12}^s$, the flavour-specific CP asymmetry $a_{fs}^s$ can be written as~\cite{Lenz:2006hd,Beneke:1998sy,Beneke:2003az}
\begin{align}\label{eq:afss-def}
a_{fs}^s \equiv {\rm Im}(\frac{\Gamma_{12}^s}{M_{12}^s})=\frac{|\Gamma_{12}^s|}{|M_{12}^s|}\sin\Phi_s\,.
\end{align}
As the semileptonic decay $B_s\to X \ell \nu_{\ell}$ is a typical example of flavour-specific decays, this asymmetry is usually also named as the semileptonic CP asymmetry~\cite{Asner:2010qj}.

Thus, in order to predict the mixing observables $\Delta M_s$, $\Delta \Gamma_s$, $\phi_s$, as well as $a_{fs}^{s}$, we need to know the off-diagonal elements $M_{12}^s$ and $\Gamma_{12}^s$ both within the SM and in the $Z^\prime$ model.

\subsection{The family non-universal $Z^\prime$ model}
\label{Sec:zprime-model}

In many well-motivated NP models with an additional ${\rm U(1)}^{\prime}$ gauge symmetry, such as string constructions and/or grand unified theories, the associated $Z^{\prime}$ boson can generally have family non-universal, flavour-changing couplings~\cite{Langacker:2008yv}. The general formalism of such a model has been detailed in Refs.~\cite{Langacker:2000ju,Barger:2009eq}.

In the physical mass-eigenstate basis, due to the non-diagonal chiral coupling matrix, the tree-level FCNC interactions appear both in the left- and in the right-handed sectors~\cite{Langacker:2000ju,Barger:2009eq}. The relevant interaction Lagrangian can be written as~\cite{Alok:2010ij}
\begin{equation}\label{eq:zprime-lag}
\mathcal{L}_{Z^{\prime}} = \frac{g}{\cos\theta_{W}}\left[(B_{sb}^{L}\,\bar{s}\gamma^{\mu}P_{L}b + B_{sb}^{R}\,\bar{s}\gamma^{\mu}P_{R}b + {\rm h.c.}) + (B_{qq}^{L}\,\bar{q}\gamma^{\mu}P_{L}\,q + B_{qq}^{R}\,\bar{q}\gamma^{\mu}P_{R}\,q)\right]Z_{\mu}^{\prime}\,,
\end{equation}
where $P_{L,R}=\frac{1}{2}(1\mp \gamma_5)$ project onto the left- and right-handed chiral fields, $g$ is the SM SU(2) coupling, and $\theta_{W}$ the weak mixing angle. For convenience, we have absorbed the ${\rm U(1)}^{\prime}$ coupling constant into the factors $B^{L,R}_{ij}$, and written the above Lagrangian in terms of the SM coupling. It should be noted that, while the flavour-changing couplings are in general complex, the diagonal ones have to be real due to the hermiticity of the Lagrangian.

Starting with the Lagrangian Eq.~(\ref{eq:zprime-lag}), and integrating out the heavy $Z^\prime$ boson, one can easily obtain the resulting effective $|\Delta B|=1$ and $|\Delta B|=2$ four-fermion interactions induced by tree-level $Z^\prime$ exchange~\cite{Langacker:2000ju,Barger:2009eq}.

\subsection{The off-diagonal element $M_{12}^s$}
\label{Sec:m12s-zprime}

The off-diagonal element $M_{12}^s$ can be decomposed into the SM and $Z^{\prime}$ parts
\begin{equation}\label{eq:m12s-def}
M_{12}^s=(M_{12}^s)_{\rm SM} + (M_{12}^s)_{Z^{\prime}}\,,
\end{equation}
where the SM contribution could be found, for example, in Refs.~\cite{Buras:1998raa,Buras:2001ra}. In our case, due to the simultaneous presence of left- and right-handed currents, the complete set of dimension-six operators consists of the following ones~\cite{Buras:2001ra,Buras:2010pz}
\begin{align}\label{eq:operator-basis}
&Q_{1}^{VLL} = (\bar{s}^{\alpha}\gamma_{\mu}P_{L}b^{\alpha})(\bar{s}^{\beta}\gamma^{\mu}P_{L}b^{\beta})\,, \quad
Q_{1}^{VRR} = (\bar{s}^{\alpha}\gamma_{\mu}P_{R}b^{\alpha})(\bar{s}^{\beta}\gamma^{\mu}P_{R}b^{\beta})\,, \nonumber\\[0.2cm]
&Q_{1}^{LR} = (\bar{s}^{\alpha}\gamma_{\mu}P_{L}b^{\alpha})(\bar{s}^{\beta}\gamma^{\mu}P_{R}b^{\beta})\,, \quad
Q_{2}^{LR} = (\bar{s}^{\alpha}P_{L}b^{\alpha})(\bar{s}^{\beta}P_{R}b^{\beta})\,,
\end{align}
where $\alpha$ and $\beta$ denote the colour indices. Because of the $V-A$ structure of $W^{\pm}$-boson exchanges, only $Q_{1}^{VLL}$ contributes within the SM. The operators $Q_{1}^{VRR}$ and $Q_{1}^{LR}$ are generated at the scale $\mu_{Z^\prime}\sim m_{Z^\prime}$, whereas $Q_{2}^{LR}$ appear through renormalization group~(RG) running to some scale different from $\mu_{Z^\prime}$.

Normalized to the effective Hamiltonian for $|\Delta B|=2$ transition within the SM~\cite{Buras:1998raa}
\begin{equation}\label{eq:Heff-sm}
\mathcal{H}_{\rm eff}^{\rm SM}(|\Delta B|=2) = \frac{G_F^2}{16\pi^2}\,m_W^2\,(V_{tb}V_{ts}^{\ast})^2\,C_1^{VLL}(\mu)\,Q_1^{VLL}(\mu) + {\rm h.c.}\,,
\end{equation}
the NP effective Hamiltonian induced by tree-level $Z^\prime$ exchange can be written as~\cite{Barger:2009eq,Buras:2010pz}
\begin{align}\label{eq:Heff-zprime}
\mathcal{H}_{\rm eff}^{Z^\prime}(|\Delta B|=2) &= \frac{G_{F}^{2}}{16\pi^{2}}\,m_{W}^{2}\,(V_{tb} V_{ts}^{*})^2\,\Big[\Delta C_{1}^{VLL}(\mu)\,Q_{1}^{VLL}(\mu) + \Delta C_{1}^{VRR}(\mu)\,Q_{1}^{VRR}(\mu)\, \nonumber \\[0.2cm]
& \quad + \Delta C_{1}^{LR}(\mu)\,Q_{1}^{LR}(\mu) + \Delta C_{2}^{LR}(\mu)\,Q_{2}^{LR}(\mu)\Big] + {\rm h.c.}\,,
\end{align}
where $G_F$ is the Fermi coupling constant, and $m_W$ denotes the $W$-boson mass. The Wilson coefficients at the high scale $\mu_{Z^{\prime}}$ are obtained by integrating out the heavy $Z^\prime$ boson, with the final results given explicitly as
\begin{align}\label{eq:deltaci-zprime}
\Delta C_{1}^{VLL}(\mu_{Z^{\prime}}) &= \frac{m_{Z}^{2}}{m_{Z^{'}}^{2}}\,\frac{64\pi^{2}}{\sqrt{2} G_{F} m_{W}^{2}}\,\frac{B^{L}_{sb}B^{L}_{sb}}{(V_{tb} V_{ts}^{*})^{2}}\,,\nonumber\\[0.2cm]
\Delta C_{1}^{VRR}(\mu_{Z^{\prime}}) &= \frac{m_{Z}^{2}}{m_{Z^{'}}^{2}}\,\frac{64\pi^{2}}{\sqrt{2} G_{F} m_{W}^{2}}\,\frac{B^{R}_{sb}B^{R}_{sb}}{(V_{tb} V_{ts}^{*})^{2}}\,, \nonumber\\[0.2cm]
\Delta C_{1}^{LR}(\mu_{Z^{\prime}}) &= \frac{m_{Z}^{2}}{m_{Z^{'}}^{2}}\,\frac{64\pi^{2}}{\sqrt{2} G_{F} m_{W}^{2}}\,\frac{2 B^{L}_{sb}B^{R}_{sb}}{(V_{tb} V_{ts}^{*})^{2}}\,, \nonumber\\[0.2cm]
\Delta C_{2}^{LR}(\mu_{Z^{\prime}}) &= 0\,,
\end{align}
where $m_{Z}$ is the $Z$-boson mass. It is noted that the last coefficient $\Delta C_{2}^{LR}$ vanishes at the high scale $\mu_{Z^{\prime}}$ in the absence of QCD effects assumed by us.

In order to include the QCD RG evolution during the calculation of $M_{12}^s$, following the method proposed in Refs.~\cite{Buras:2001ra,Gorbahn:2009pp}, we shall evaluate the Wilson coefficients and the hadronic matrix elements of local operators at the high scale $\mu_{Z^\prime}$. The final $Z^\prime$-boson contribution to the element $M_{12}^s$ can be written as
\begin{align} \label{eq:m12s-zprime}
(M_{12}^s)_{Z^{\prime}} &= \frac{1}{2m_{B_s}}\,\langle B_s^0|\mathcal{H}_{\rm eff}^{\rm Z^\prime}(\Delta B=2)|\bar B_s^0\rangle\, \nonumber\\[0.2cm]
&= \frac{G_{F}^{2}}{48\pi^{2}}\,m_{W}^{2}\,(V_{tb}V_{ts}^{*})^2\,m_{B_s}\,f_{B_s}^2\,\Big[(\Delta C_{1}^{VLL}(\mu_{Z^\prime}) + \Delta C_{1}^{VRR}(\mu_{Z^\prime}))\,P_{1}^{VLL}(\mu_{Z^\prime})\, \nonumber\\[0.2cm]
& \qquad \qquad + \Delta C_{1}^{LR}(\mu_{Z^\prime})\,P_{1}^{LR}(\mu_{Z^\prime}) \Big]\,,
\end{align}
where $m_{B_s}$ and $f_{B_s}$ are the mass and decay constant of $B_s$ meson, respectively. The coefficients $P_i^{a}(\mu)$ collect compactly all RG effects from scales below $\mu_{Z^\prime}$ and hadronic matrix elements obtained by lattice methods at low scales, and are defined as~\cite{Buras:2001ra}
\begin{equation}
\langle B_s^0|Q_i^a(\mu)|\bar B_s^0\rangle = \frac{2}{3}m_{B_s}^2 f_{B_s}^2 P_i^a(\mu)\,.
\end{equation}
Analytic formulae for $P_i^{a}(\mu)$ given explicitly in terms of the QCD RG factors and the non-perturbative bag parameters could be found in Ref.~\cite{Buras:2001ra}.

\subsection{The off-diagonal element $\Gamma_{12}^s$}
\label{Sec:gamma12s-zprime}

The off-diagonal element $\Gamma_{12}^s$, being an inclusive quantity stemming from decays into final states common to $B_s^0$ and $\bar B_s^0$ mesons, could be computed using the heavy quark expansion~(HQE)~\cite{Shifman:1984wx,Bigi:1992su}, which is a simultaneous expansion in $\Lambda_{\rm QCD}/m_b$ and $\alpha_s(m_b)$. It is now known to next-to-leading-order~(NLO) both in $\Lambda_{\rm QCD}/m_b$~\cite{Beneke:1996gn,Dighe:2001gc} and in $\alpha_s(m_b)$~\cite{Beneke:1998sy,Beneke:2003az,Ciuchini:2003ww}, with a recent theoretical update made in Ref.~\cite{Lenz:2006hd}. Motivated by the like-sign dimuon charge asymmetry observed by the D0 collaboration~\cite{Abazov:2010hv,Abazov:2011yk}, the possibility to have sizable NP effects in $\Gamma_{12}^s$ has been discussed in the literature~\cite{BsBsbarmixing-new,Bobeth:2011st,Alok:2010ij,dimuoncharge}.

Following the notation specified by Beneke {\it et al.}~\cite{Beneke:1998sy,Beneke:1996gn}, the off-diagonal element $\Gamma_{12}^s$, which is related via optical theorem to the absorptive part of the forward-scattering amplitude, can be written as
\begin{equation}\label{eq:gamma12-def}
\Gamma_{12}^{s} = \frac{1}{2 m_{B_s}}\,\langle B_s^0|\mathcal{T}|\bar{B}_s^0\rangle\,,
\end{equation}
with the transition operator $\mathcal{T}$ defined by
\begin{equation}\label{eq:T-def}
\mathcal{T} = {\rm Im}\,i\int d^{4}x\,T[\mathcal{H}_{\rm eff}(x)\,\mathcal{H}_{\rm eff}(0)]\,.
\end{equation}
Here $\mathcal{H}_{\rm eff}$ is the low-energy effective Hamiltonian for $|\Delta B|=1$ transition, which is obtained by integrating out the heavy particles above the scale $\mu_b\simeq m_b$, such as the top quark, $W$ and $Z$ bosons within the SM, as well as the heavy $Z^\prime$ boson in our case. It can be generally decomposed as
\begin{equation}
\mathcal{H}_{\rm eff} = \mathcal{H}_{\rm eff}^{\rm SM}(|\Delta B|=1) + \mathcal{H}_{\rm eff}^{Z^\prime}(|\Delta B|=1)\,,
\end{equation}
where a detailed review and explicit expressions of the SM part $\mathcal{H}_{\rm eff}^{\rm SM}(|\Delta B|=1)$ could be found, for example, in Ref.~\cite{Buras:1998raa}.

The $Z^\prime$ contribution to $\Gamma_{12}^s$ arises from dimension-six operators of the form $b\to s f\bar f$~(with $f$ denoting a light fermion), most of which are, however, constrained to be small by the rare B-meson decays. For example, the semileptonic operators $b\to s e^+e^-$, $b\to s \mu^+\mu^-$ are severely constrained by the $b\to s \ell^+ \ell^-$ processes~\cite{rareB}, while the light-quark operators $b\to s u\bar u$, $b\to s d\bar d$, $b\to s s\bar s$ are strongly bounded by the various hadronic B-meson decays like $B\to \pi K$ and $B\to \phi K$~\cite{hadronicB}. Furthermore, it is found that the contribution from $b\to s \tau^+ \tau^-$ clearly fail to describe the $B_s-\bar B_s$ mixing data within $68\%$ C.L.~\cite{Bobeth:2011st}. Thus, in this paper we shall only consider the $Z^\prime$ contribution coming from the four-quark operator $b\to s c\bar c$.

Starting from the general couplings of Eq.~(\ref{eq:zprime-lag}), we get four types of operators, $(\bar s b)_{V-A}(\bar c c)_{V \pm A}$ and $(\bar s b)_{V+A}(\bar c c)_{V \pm A}$, where the former two are already present in the SM. Since the chirality of the flavour-changing part for the latter two operators is flipped, to run their Wilson coefficients down to the lower scale $\mu_b\sim m_b$, we have to calculate the corresponding anomalous dimensional matrix. Furthermore, due to the presence of these new operator structures beyond the SM, the calculation of their contributions to $\Gamma_{12}^s$ needs also to be extended. Since both of these calculations are non-trivial, for simplicity we shall restrict ourselves to the case where only purely left-handed flavour-changing $B_{sb}^L$ coupling is present, which is usually assumed in the literature~\cite{BsBsbarmixing-new,Bobeth:2011st,Alok:2010ij}. For the flavour-diagonal part, on the other hand, both the left- and right-handed CKM-favored couplings $B_{cc}^L$ and $B_{cc}^R$ are taken into account.

With these assumptions, the final $Z^\prime$-boson contribution to the off-diagonal element $\Gamma_{12}^s$ can be written as~\cite{Barger:2009eq}
\begin{align} \label{eq:Heff-DeltaB1-zprime}
\mathcal{H}_{\rm eff}^{Z^\prime}(|\Delta B|=1) &= \frac{G_{F}}{\sqrt{2}}\, \frac{2m_{Z}^{2}}{m_{Z^\prime}^{2}}\,\left[B^{L}_{sb}\,B^{L}_{cc}\, (\bar{s}b)_{V-A}(\bar{c}c)_{V-A} + B^{L}_{sb}\,B^{R}_{cc}\,(\bar{s}b)_{V-A}(\bar{c}c)_{V+A}\right]\, \nonumber \\[0.2cm]
& = -\frac{G_{F}}{\sqrt{2}}\,V_{tb}V_{ts}^{\ast}\,\left[\Delta C_3\,Q_3 + \Delta C_5\,Q_5\right]\,,
\end{align}
where $(\bar{q}q^\prime)_{V \pm A}=\bar{q}\gamma_{\mu}(1 \pm \gamma_5)q^\prime$, and in the second line we have normalized the result to the SM effective Hamiltonian~\cite{Buras:1998raa}. Thus, with our assumptions, the $Z^\prime$ effects can be understood as corrections to the Wilson coefficients of SM QCD-penguin operators $Q_3$ and $Q_5$. The new Wilson coefficients at the high scale $\mu_{Z^\prime}$ are given, respectively, as
\begin{align} \label{eq:DeltaC3C5-zprime}
\Delta C_{3}(\mu_{Z^{\prime}}) = - \frac{2}{V_{tb} V_{ts}^{\ast}}\,\frac{m_{Z}^{2}}{m_{Z^{\prime}}^{2}}\, B^{L}_{sb} B^{L}_{cc}\,, \qquad
\Delta C_{5}(\mu_{Z^{\prime}}) = - \frac{2}{V_{tb} V_{ts}^{\ast}}\,\frac{m_{Z}^{2}}{m_{Z^{\prime}}^{2}}\, B^{L}_{sb} B^{R}_{cc}\,.
\end{align}
Since no new operator structures beyond the SM appear in our case, the remaining calculation of the element $\Gamma_{12}^s$ is the same as that within the SM, the details of which could be found in Refs.~\cite{Lenz:2006hd,Beneke:1998sy,Beneke:2003az,Beneke:1996gn}.

In the presence of the effective Hamiltonian Eq.~(\ref{eq:Heff-DeltaB1-zprime}), the indirect CP asymmetry in $\bar{B}_d\to J/\psi K_S$ is also affected due to the NP contribution to the decay $b\to s c\bar{c}$, and the effective measured $\sin2\beta^{\rm meas}$ is given by~\cite{Alok:2010ij,Chiang:2009ev}
\begin{align}
\sin2\beta^{\rm meas}=\sin2\beta +2 |r|\cos2\beta\sin\phi\cos\delta^a\,,
\end{align}
where $\phi$ and $\delta^a$ are the NP weak and the SM strong phase, respectively. The parameter $|r|$ denotes the magnitude of the ratio between NP and SM $b\to s c\bar{c}$ decay amplitudes. In terms of the $Z^\prime$ couplings, it is given explicitly as
\begin{align}
|r|=\biggl|\frac{2m_Z^2}{m_{Z^\prime}^2\,V_{cb}V_{cs}^{\ast}\,a_2}\,B_{sb}^L\,(B_{cc}^L+B_{cc}^R)\biggr|\,,
\end{align}
where $a_2\simeq 0.17$ is the effective Wilson coefficient within the SM. Numerically we shall use the estimation $|r|\leq 26.5\%$ at $1\sigma$~\cite{Alok:2010ij} as a constraint on the model parameter space.

For convenience, when performing the RG running between the scales $\mu_{Z^\prime}$ and $\mu_b$, we shall neglect the small RG effects in the six-flavour theory throughout the paper. It should be noted that all the above discussions apply also to the $B_d^0-\bar B_d^0$ mixing, with the replacement $s\to d$.

\section{Numerical results and discussions}
\label{Sec:numericalresult}

Equipped with the formalism discussed in previous section, as well as the theoretical input parameters listed in the appendix, we proceed to present our numerical results and discussions in this section.

\subsection{The SM predictions and the experimental data}

\begin{table}[t]
\begin{center}
\caption{\label{tab:numerical-result} \small Our theoretical predictions for the $B_s^0-\bar B_s^0$ mixing observables within the SM, with the corresponding experimental data given in the second column. For convenience, the SM predictions for some relevant $B_d^0-\bar B_d^0$ mixing observables are also given.}
\vspace{0.2cm}
\doublerulesep 0.8pt \tabcolsep 0.30in
\begin{tabular}{l c c}
\hline\hline
observable & experimental data & SM prediction \\
\hline
$\Delta M_{s}~{\rm [ps^{-1}]}$ & $17.77\pm 0.12$~\cite{Abulencia:2006mq} & $17.27^{+2.70}_{-2.55}$       \\
                               & $17.73\pm 0.05$~\cite{LHCbnote50}       & \\[0.1cm]
$\phi_{s}~{\rm [rad]}$ & $\in[-1.04,-0.04] \cup [-3.10,-2.16]$~\cite{CDF:2011af} & $-0.0421^{+0.0054}_{-0.0054}$ \\
                       & $0.07\pm 0.18$~\cite{LHCb:2011ab}                       & \\[0.1cm]
$\Delta\Gamma_{s}~{\rm [ps^{-1}]}$ & $0.075\pm 0.036$~\cite{CDF:2011af} & $0.065^{+0.024}_{-0.028}$ \\
                                   & $0.123\pm 0.031$~\cite{LHCb:2011ab}  & \\[0.1cm]
$a_{fs}^{s}~{\rm [\%]}$ & $-1.81\pm 1.06$~\cite{Abazov:2011yk} & $0.0033^{+0.0007}_{-0.0010}$ \\[0.1cm]
\hline
$\Delta M_{d}~{\rm [ps^{-1}]}$ & $0.507\pm 0.004$~\cite{Asner:2010qj} & $0.534^{+0.117}_{-0.110}$ \\[0.1cm]
$\phi_{d}~{\rm [degree]}$ & $42.8\pm 1.6$~\cite{Asner:2010qj} & $48.57^{+6.44}_{-6.42}$ \\
\hline \hline
\end{tabular}
\end{center}
\end{table}

Within the SM, our predictions for the $B_s^0-\bar B_s^0$ mixing observables are listed in Table~\ref{tab:numerical-result}. The theoretical uncertainties are obtained by varying each input parameter within its respective range specified in the appendix and adding the individual uncertainty in quadrature. The corresponding experimental data, which has been summarized in the introduction section, is collected in the second column. For convenience, we also give in Table~\ref{tab:numerical-result} some relevant mixing observables in the $B_d$-meson system that will be used later.

As can be seen from Table~\ref{tab:numerical-result}, our result for the mass difference $\Delta M_s$ agrees quite well the experimental data, which is therefore expected to give a strong constraint on the NP parameter space. For the CP-violating phase $\phi_s$, while the SM prediction is quite precise, the experimental data still has a large uncertainty, with the central value deviating from the SM result. Our prediction for the width difference $\Delta \Gamma_s$, while being consistent with the CDF measurement, is about $1\sigma$ below the LHCb measurement. There is, however, a tension for the flavour-specific CP asymmetry $a_{fs}^s$ between the SM prediction and the D0 measurement, which has triggered a lot of investigations both within the SM and in various NP models~\cite{BsBsbarmixing-new,Bobeth:2011st,Alok:2010ij,dimuoncharge}. Thus, it is interesting to check if the parameter space allowed by $\Delta M_s$ could simultaneously explain the remaining observables, especially $\Delta \Gamma_s$ and $a_{fs}^s$ in a specific NP model.

In the following, we shall revisit the $B_s^0-\bar B_s^0$ mixing in a family non-universal $Z^\prime$ model. We shall use two different data sets: the first one~(D1) corresponds to the first row and the second~(D2) the second row listed in Table~\ref{tab:numerical-result}, while for $a_{fs}^s$ we use the latest D0 measurement for both sets. In addition, for both data sets, our analyses are further divided into the following two scenarios: $(M_{12}^{s})_{\rm Z^\prime}\neq 0$, $(\Gamma_{12}^{s})_{\rm Z^\prime}=0$~(named as S1) and $(M_{12}^{s})_{\rm Z^\prime}\neq 0$, $(\Gamma_{12}^{s})_{\rm Z^\prime}\neq 0$~(named as S2). To obtain the allowed parameter space, we shall use the theoretical predictions with $1\sigma$, as well as the experimental data with $2\sigma$ error bars as constraints\footnote{For the CP-violating phase $\phi_s$, on the other hand, in the data set D1 we still use the $68\%$ C.L. region measured by the CDF collaboration~\cite{CDF:2011af}. }.

\subsection{$Z^\prime$ effects in scenario S1}

In this subsection, we consider the first scenario S1 with $(M_{12}^{s})_{\rm Z^\prime}\neq 0$ and $(\Gamma_{12}^{s})_{\rm Z^\prime}=0$, where all the $Z^\prime$ effects are encoded in the flavour-changing couplings $B_{sb}^{L}=|B_{sb}^{L}|e^{i\phi_{sb}^L}$ and $B_{sb}^{R}=|B_{sb}^{R}|e^{i\phi_{sb}^R}$. In terms of the four parameters $|B_{sb}^{L,R}|$ and $\phi_{sb}^{L,R}$, and fixing the $Z^\prime$-boson mass at $m_{Z^\prime}=400~{\rm GeV}$, the off-diagonal element $M_{12}^s$ can be formally written as
\begin{align}\label{eq:m12s-numerical}
M_{12}^s = (M_{12}^s)_{\rm SM}\biggl[1 & + 2.11\,|B_{sb}^{L}[10^{-2}]|^2\,e^{2i(\phi_{sb}^L + 1.21^{\circ})} + 2.11\,|B_{sb}^{R}[10^{-2}]|^2\,e^{2i(\phi_{sb}^R + 1.21^{\circ})}\, \nonumber \\[0.2cm]
& - 16.41\,|B_{sb}^{L}[10^{-2}]|\,|B_{sb}^{R}[10^{-2}]|\,e^{2i(\phi_{sb}^L/2 + \phi_{sb}^R/2 + 1.21^{\circ})} \biggr]\,,
\end{align}
where the magnitudes of the couplings $|B_{sb}^{L,R}|$ are given in unit of $10^{-2}$. From Eq.~(\ref{eq:m12s-numerical}), one can see that the fourth term has only a marginal effect if $|B_{sb}^{L}| \gg |B_{sb}^{R}|$ or $|B_{sb}^{L}| \ll |B_{sb}^{R}|$, but gives the dominant contribution if $|B_{sb}^{L}| \simeq |B_{sb}^{R}|$. It is also noted that the $Z^\prime$-boson contribution with a single left-handed coupling~(the second term) is the same as that with a single right-handed coupling~(the third term).

Thus, based on the above observations, we shall consider the following two limiting cases: in case~I $B_{sb}^{L}$ varies arbitrary and $B_{sb}^{R}=0$, while in case~II the simplification $B_{sb}^{L}=B_{sb}^{R}$ is assumed.

\subsubsection{Case~I with $B_{sb}^{L}$ arbitrary and $B_{sb}^{R}=0$}

\begin{figure}[htbp]
\centering
\includegraphics[width=15cm]{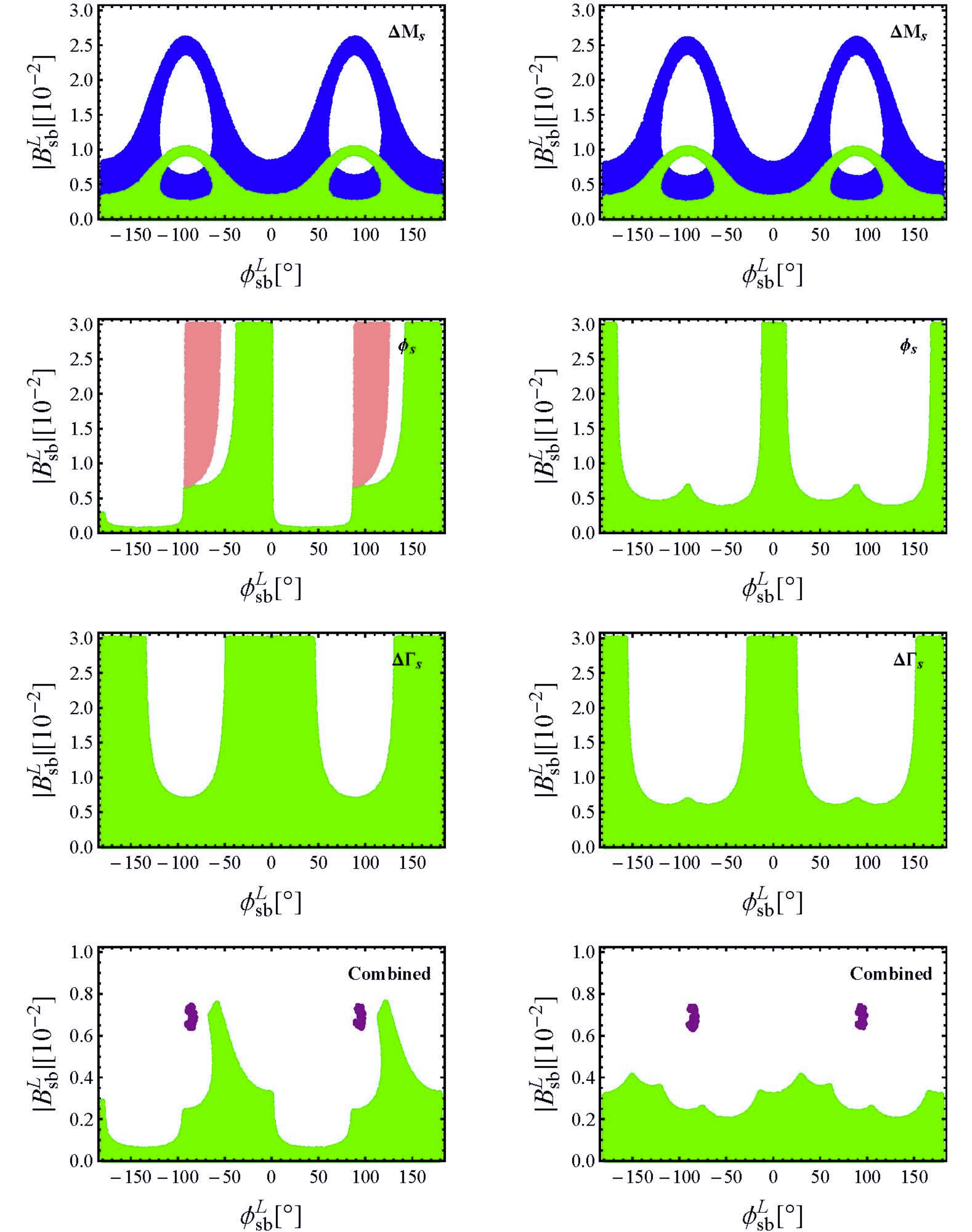}
\caption{\label{fig:Bs-BsbL-final} \small The $\phi_{sb}^{L}$-$|B_{sb}^{L}|$ parameter space in case~I, allowed by the $B_s^0-\bar B_s^0$ mixing observables. The left and the right panel correspond to the data sets D1 and D2, respectively. The green and the blue areas correspond to the case by choosing a $Z^\prime$ boson with $m_{Z^\prime}=400~{\rm GeV}$ and $m_{Z^\prime}=1~{\rm TeV}$, respectively, while the purple area is allowed by the measured $a_{fs}^s$ with $1\sigma$ error bars. The second solution for $\phi_s$, $\phi_s\in[-3.10,-2.16]~{\rm rad}$~\cite{CDF:2011af}, is denoted by the pink area. See text for details.}
\end{figure}

In this case, the $Z^\prime$ effect is parameterized by the two parameters $|B_{sb}^{L}|$ and $\phi_{sb}^{L}$, which could be severely constrained by the measured observables $\Delta M_s$, $\phi_s$, $\Delta \Gamma_s$ and $a_{fs}^s$. This is clearly shown in Fig.~\ref{fig:Bs-BsbL-final}, where the left and the right panel correspond to the data sets D1 and D2, respectively. The green and the blue areas correspond to the case by choosing a $Z^\prime$ boson with $m_{Z^\prime}=400~{\rm GeV}$ and $m_{Z^\prime}=1~{\rm TeV}$, respectively, while the purple area is allowed by the measured $a_{fs}^s$ with $1\sigma$ error bars. The second solution for $\phi_s$, $\phi_s\in[-3.10,-2.16]~{\rm rad}$~\cite{CDF:2011af}, is denoted by the pink area.

From Fig.~\ref{fig:Bs-BsbL-final}, we make the following observations:
\begin{enumerate}

\item[$\bullet$] as shown in the first two plots, decreasing the $Z^\prime$-boson mass reduces the allowed parameter space, which is also true for the other observables. Thus, for simplicity, in the following discussions, we shall choose a fixed $Z^\prime$-boson mass with $m_{Z^\prime}=400~{\rm GeV}$.

\item[$\bullet$] these mixing observables have a different and complementary dependence on the $Z^\prime$ coupling $B_{sb}^L$, and their combined constraints are very strong, as shown in the last two plots, where the parameter space allowed by the individual observable is reduced significantly.

\item[$\bullet$] the pink area in the third plot that is allowed by the second solution for the CP-violating phase $\phi_s$, $\phi_s\in[-3.10,-2.16]~{\rm rad}$~\cite{CDF:2011af}, is already excluded, once constraints from the measured $\Delta M_s$ and $\Delta \Gamma_s$ are taken into account.

\item[$\bullet$] comparing the left and the right panel, the data set D2 gives a more stringent constraint on the magnitude $|B_{sb}^{L}|$. However, at present no constraint on the weak phase $\phi_{sb}^{L}$ could be obtained for both data sets from the measured observables $\Delta M_s$, $\phi_s$ and $\Delta \Gamma_s$.

\item[$\bullet$] as shown in the last two plots, there is no overlap between the parameter space allowed by $a_{fs}^s$ with $1\sigma$ error bars, as well as the one allowed simultaneously by the three observables $\Delta M_s$, $\phi_s$ and $\Delta \Gamma_s$. Thus, within such a specific $Z^\prime$ model, it is difficult to explain the observed flavour-specific CP asymmetry $a_{fs}^s$ by the D0 collaboration~\cite{Abazov:2011yk}.

\end{enumerate}

\subsubsection{Case~II with the simplification $B_{sb}^{L}=B_{sb}^{R}$}

In this case, due to the simultaneous presence of left- and right-handed currents, the $Z^\prime$-boson contribution to $M_{12}^s$ is dominated through the generation of the LR operators that renormalize strongly under QCD~\cite{Buras:2001ra,Buras:2010pz}. With the simplification $B_{sb}^{L}=B_{sb}^{R}$, contributions from the VLL and VRR operators are completely canceled by that from the LR operators. This can be clearly seen from the numerical result given in Eq.~(\ref{eq:m12s-numerical}).

\begin{figure}[t]
\centering
\includegraphics[width=15cm]{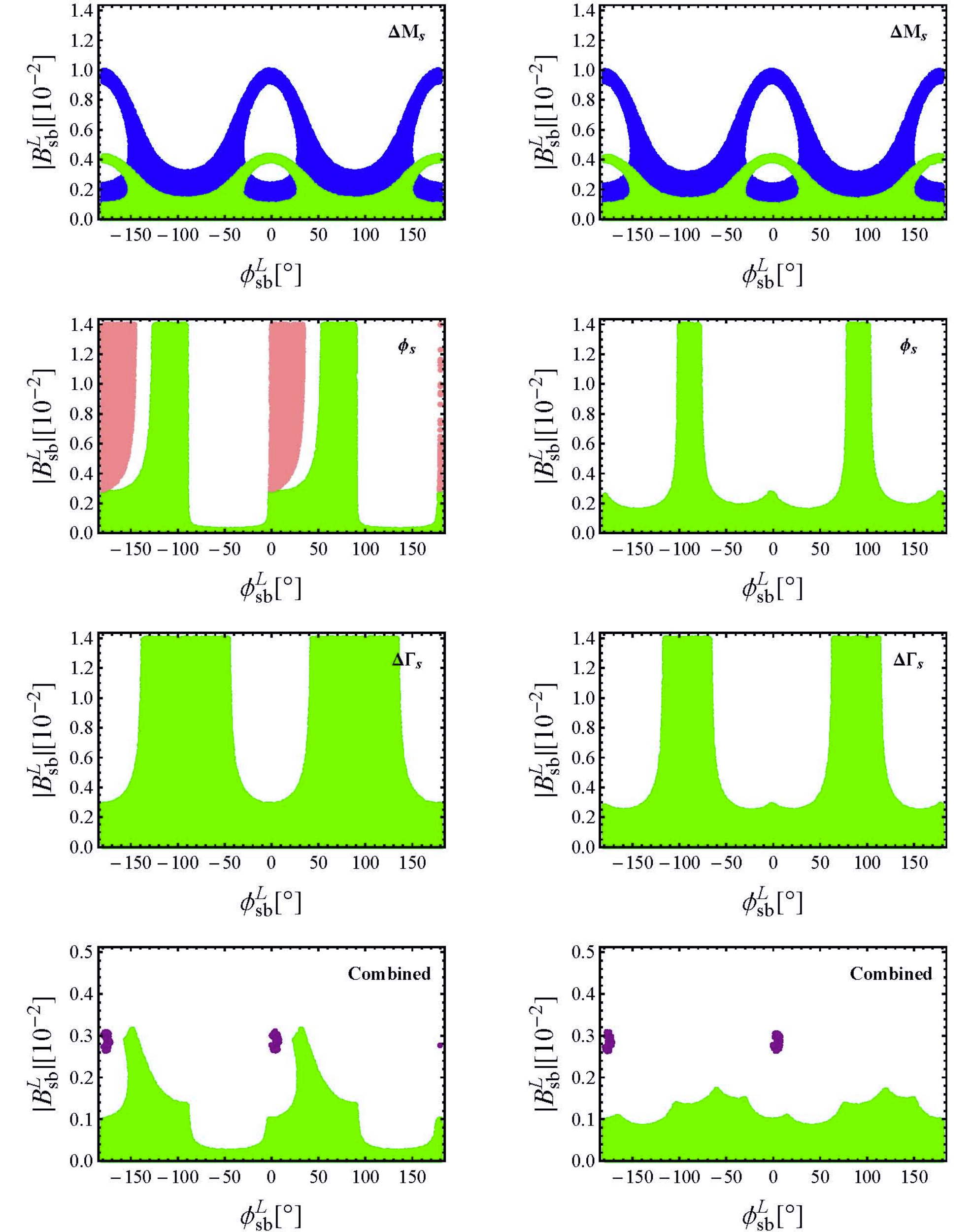}
\caption{\label{fig:Bs-BsbLR-final} \small The $\phi_{sb}^{L}$-$|B_{sb}^{L}|$ parameter space in case~II, allowed by the measured $B_s^0-\bar B_s^0$ mixing observables. The other captions are the same as in Fig.~\ref{fig:Bs-BsbL-final}.}
\end{figure}

Corresponding to case~II, we show in Fig.~\ref{fig:Bs-BsbLR-final} the $\phi_{sb}^{L}$-$|B_{sb}^{L}|$ parameter space allowed by the measured $B_s^0-\bar B_s^0$ mixing observables. The other captions are the same as in Fig.~\ref{fig:Bs-BsbL-final}. In this case, the following observations could be made:
\begin{enumerate}

\item[$\bullet$] due to the presence of the LR operators, the mixing observables show a quite different dependence on the NP weak phase $\phi_{sb}^L$ than in case~I. The allowed ranges for the magnitude $|B_{sb}^L|$ are significantly reduced, being about half of that obtained in case~I.

\item[$\bullet$] as in case~I, with constraints from the measured $\Delta M_s$, $\phi_s$ and $\Delta \Gamma_s$ taken into account, we still could not reproduce the present measured $1\sigma$ experimental ranges of $a_{fs}^s$ by the D0 collaboration~\cite{Abazov:2011yk}.

\item[$\bullet$] in both case~I and case~II, the mixing observables $\phi_s$, $\Delta \Gamma_s$ and $a_{fs}^s$, which are all related to the phase of $M_{12}^s$, show a strong dependence on the NP weak phase $\phi_{sb}^L$.

\end{enumerate}

Thus, combining the above two limiting cases, it is concluded that the scenario S1 could not explain the measured like-sign dimuon charge asymmetry by the D0 collaboration~\cite{Abazov:2011yk}, once constraints from $\Delta M_s$, $\phi_s$ and $\Delta \Gamma_s$ are taken into account. This motivates us to consider the second scenario S2 with $(M_{12}^{s})_{\rm Z^\prime}\neq 0$ and $(\Gamma_{12}^{s})_{\rm Z^\prime}\neq 0$.

\subsection{$Z^\prime$ effects in scenario S2}

In the scenario S2, the $Z^\prime$ boson contributes both to $M_{12}^s$ and to $\Gamma_{12}^s$. With our assumptions discussed in Sec.~\ref{Sec:gamma12s-zprime}, the $Z^\prime$ effects are now parameterized by the four parameters $|B_{sb}^L|$, $\phi_{sb}^L$, $B_{cc}^L$ and $B_{cc}^R$. In terms of them, the off-diagonal element $\Gamma_{12}^s$ can be formally written as
\begin{align}\label{eq:g12s-numerical}
\Gamma_{12}^s = (\Gamma_{12}^s)_{\rm SM}\biggl[1 - 0.033\,|B_{sb}^{L}[10^{-2}]|\,B_{cc}^{L}\,e^{i(\phi_{sb}^L + 1.77^{\circ})} + 0.010\,|B_{sb}^{L}[10^{-2}]|\,B_{cc}^{R}\,e^{i(\phi_{sb}^L + 3.28^{\circ})} \biggr]\,,
\end{align}
where the $Z^\prime$-boson mass is fixed at $m_{Z^\prime}=400~{\rm GeV}$. One can see that, depending on the relative strength of the couplings $B_{cc}^{L}$ and $B_{cc}^{R}$, as well as the weak phase $\phi_{sb}^L$, the $Z^\prime$ boson could contribute either constructively or destructively to the off-diagonal element $\Gamma_{12}^s$. Since the flavour-changing couplings $|B_{sb}^L|$ and $\phi_{sb}^L$ could be constrained by the measured observables $\Delta M_s$ and $\phi_s$, we shall focus on the flavour-diagonal couplings $B_{cc}^L$ and $B_{cc}^R$, which could be bounded by the indirect CP asymmetry in $\bar{B}_d\to J/\psi K_S$~\cite{Alok:2010ij,Chiang:2009ev}.

\begin{figure}[htbp]
\centering
\includegraphics[width=15cm]{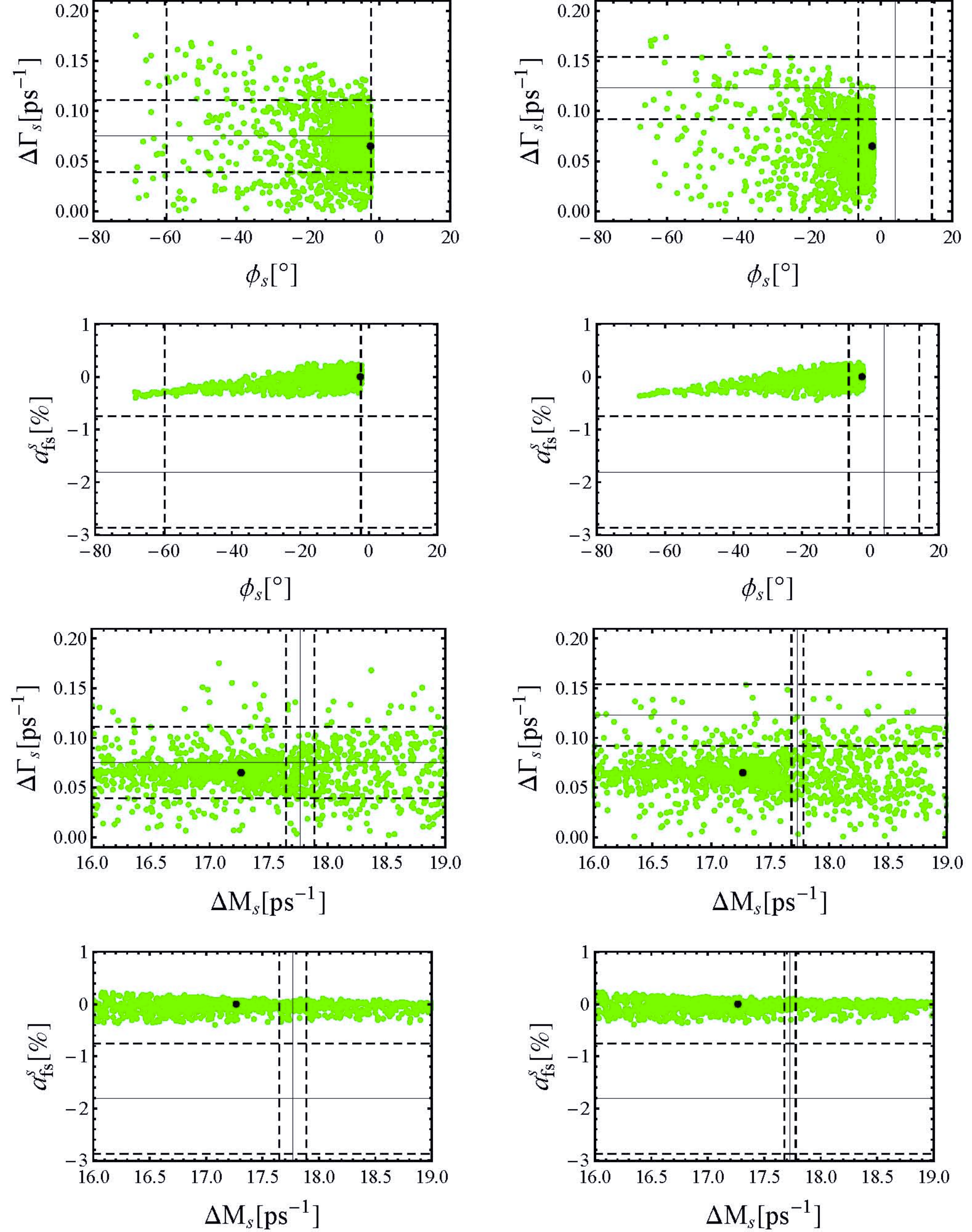}
\caption{\label{fig:Bs-BsbLBccL-final} \small The correlation plot between $\Delta \Gamma_s$, $a_{fs}^s$ and $\phi_s$, $\Delta M_s$ for Case~I where only the couplings $B_{sb}^L$ and $B_{cc}^L$ are present. The horizontal and vertical lines correspond to the experimental data with $1\sigma$ error bars, while the SM prediction is indicated as a black point. The left and the right panel correspond to the data sets D1 and D2, respectively. The $Z^\prime$-boson mass is fixed at $m_{Z^\prime}=400~{\rm GeV}$. See text for details.}
\end{figure}

\begin{figure}[t]
\centering
\includegraphics[width=15cm]{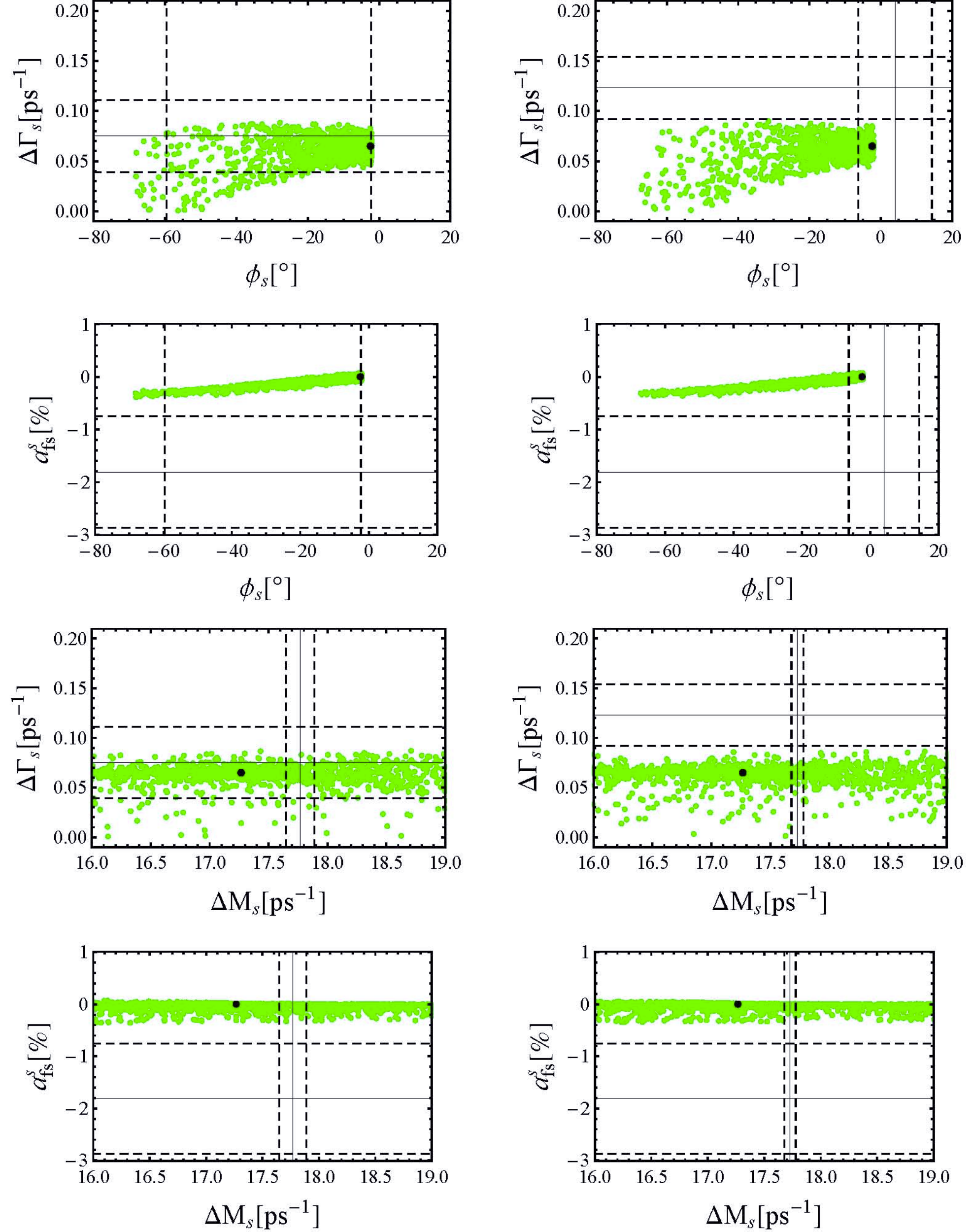}
\caption{\label{fig:Bs-BsbLBccR-final} \small The same as in Fig.~\ref{fig:Bs-BsbLBccL-final} but for Case~II where only the couplings $B_{sb}^L$ and $B_{cc}^R$ are present. The other captions are the same as in Fig.~\ref{fig:Bs-BsbLBccL-final}.}
\end{figure}

\begin{figure}[t]
\centering
\includegraphics[width=15cm]{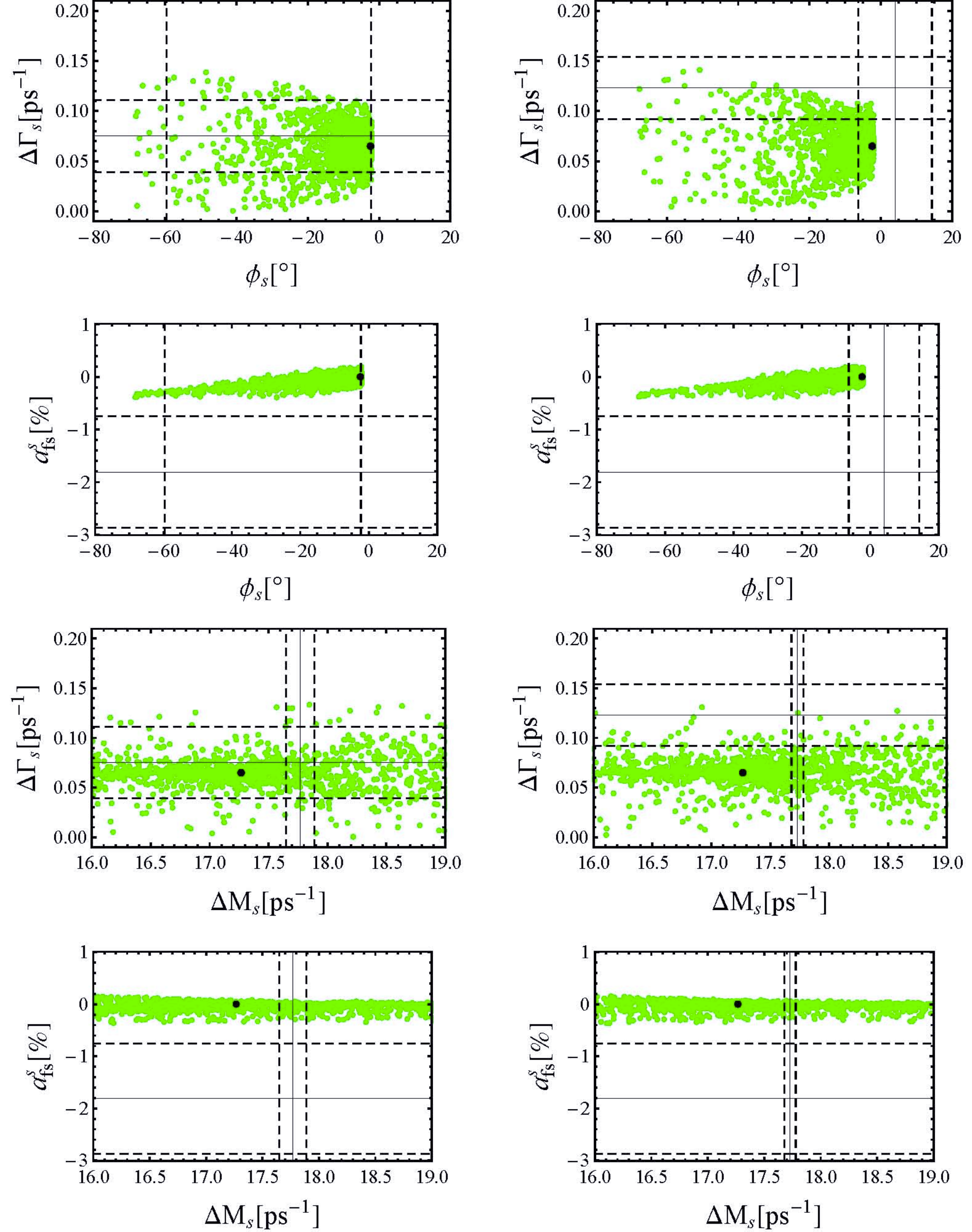}
\caption{\label{fig:Bs-BsbLBccLR-final} \small The same as in Fig.~\ref{fig:Bs-BsbLBccL-final} but for Case~III where both the couplings $B_{sb}^L$ and $B_{cc}^{L,R}$ are present, with the simplification $B_{cc}^L=B_{cc}^R$. The other captions are the same as in Fig.~\ref{fig:Bs-BsbLBccL-final}.}
\end{figure}

Thus, in this subsection, imposing the constraints from $\Delta M_s$, $\phi_s$, as well as the indirect CP asymmetry in $\bar{B}_d\to J/\psi K_S$~\cite{Alok:2010ij,Chiang:2009ev}, we shall investigate if the specific $Z^\prime$ model could explain the measured $\Delta \Gamma_s$ and $a_{fs}^s$. For simplicity, with respect to the flavour-diagonal couplings $B_{cc}^L$ and $B_{cc}^R$, our analyses are further divided into the following three different cases:
\begin{enumerate}

\item[$\bullet$] Case~I with $B_{cc}^L$ arbitrary and $B_{cc}^R=0$.

\item[$\bullet$] Case~II with $B_{cc}^R$ arbitrary and $B_{cc}^L=0$.

\item[$\bullet$] Case~III with the simplification $B_{cc}^L=B_{cc}^R$.

\end{enumerate}
With the above assumptions, the $Z^\prime$ effects can be totally parameterized by the three parameters $|B_{sb}^L|$, $\phi_{sb}^L$ and $B_{cc}^L$~(or $B_{cc}^R$). With the constraints from $\Delta M_s$, $\phi_s$ and the indirect CP asymmetry in $\bar{B}_d\to J/\psi K_S$ imposed, our predictions for $\Delta \Gamma_s$ and $a_{fs}^s$ in such a $Z^\prime$ model are shown in Figs.~\ref{fig:Bs-BsbLBccL-final}, \ref{fig:Bs-BsbLBccR-final} and \ref{fig:Bs-BsbLBccLR-final}, corresponding respectively to the three different cases listed above. In these figures, the horizontal and vertical lines correspond to the experimental data with $1\sigma$ error bars, while the SM prediction is indicated as a black point. The left and the right panel correspond to the data sets D1 and D2, respectively. Here we have fixed the $Z^\prime$-boson mass at $m_{Z^\prime}=400~{\rm GeV}$.

From the correlation plots shown in Figs.~\ref{fig:Bs-BsbLBccL-final}, \ref{fig:Bs-BsbLBccR-final} and \ref{fig:Bs-BsbLBccLR-final}, the following observations are made:
\begin{enumerate}

\item[$\bullet$] in all these three different cases, once constraints from $\Delta M_s$, $\phi_s$ and the indirect CP asymmetry in $\bar{B}_d\to J/\psi K_S$ are imposed, we still could not explain the measured like-sign dimuon charge asymmetry by the D0 collaboration~\cite{Abazov:2011yk}, at least within its $1\sigma$ ranges. This is due to the fact that the flavour-diagonal couplings $B_{cc}^{L}$ and $B_{cc}^R$ are already severely constrained by the indirect CP asymmetry in $\bar{B}_d\to J/\psi K_S$.

\item[$\bullet$] for the data set D1, the predicted values of $\Delta \Gamma_s$ in these three different cases are all compatible with the experimental data within $1\sigma$ error bars. However, for the data set D2, the result obtained in Case~II is below the present measured $1\sigma$ experimental ranges by the LHCb collaboration, with the other two cases being consistent with the data.

\item[$\bullet$] the flavour-specific CP asymmetry $a_{fs}^s$ shows a strong correlation with the observables $\Delta M_s$ and $\phi_s$, and it is therefore expected that more stringent information about the former could be obtained from the latter two observables, all of which will be measured more precisely by the LHCb collaboration in the near future.

\end{enumerate}

Thus, even with a contribution from tree-level $Z^\prime$-induced $b\to c\bar{c}s$ operators included, a large correction to $\Gamma_{12}^s$ in such a specific $Z^\prime$ model is excluded. It is therefore impossible to give a simultaneous explanation for the latest measurements of $B_s^0-\bar B_s^0$ mixing observables, especially for the like-sign dimuon charge asymmetry observed by the D0 collaboration~\cite{Abazov:2011yk}.

\subsection{$Z^\prime$ effects on $B_d^0-\bar B_d^0$ mixing}
\label{Sec:Bd-mixing}

For completeness, in this subsection we discuss the $Z^\prime$ effects on the $B_d^0-\bar B_d^0$ mixing observables $\Delta M_d$ and $\phi_d$, assuming that the NP contribution comes only from the flavour-changing $Z^\prime$ couplings $B_{db}^{L,R}$. As can be seen from Table~\ref{tab:numerical-result}, our SM predictions and the experimental data for these two observables agree quite well with each other. It is therefore expected that these measurements could exert strong constraints on the $Z^\prime$ parameter space.

\begin{figure}[t]
\centering
\includegraphics[width=15cm]{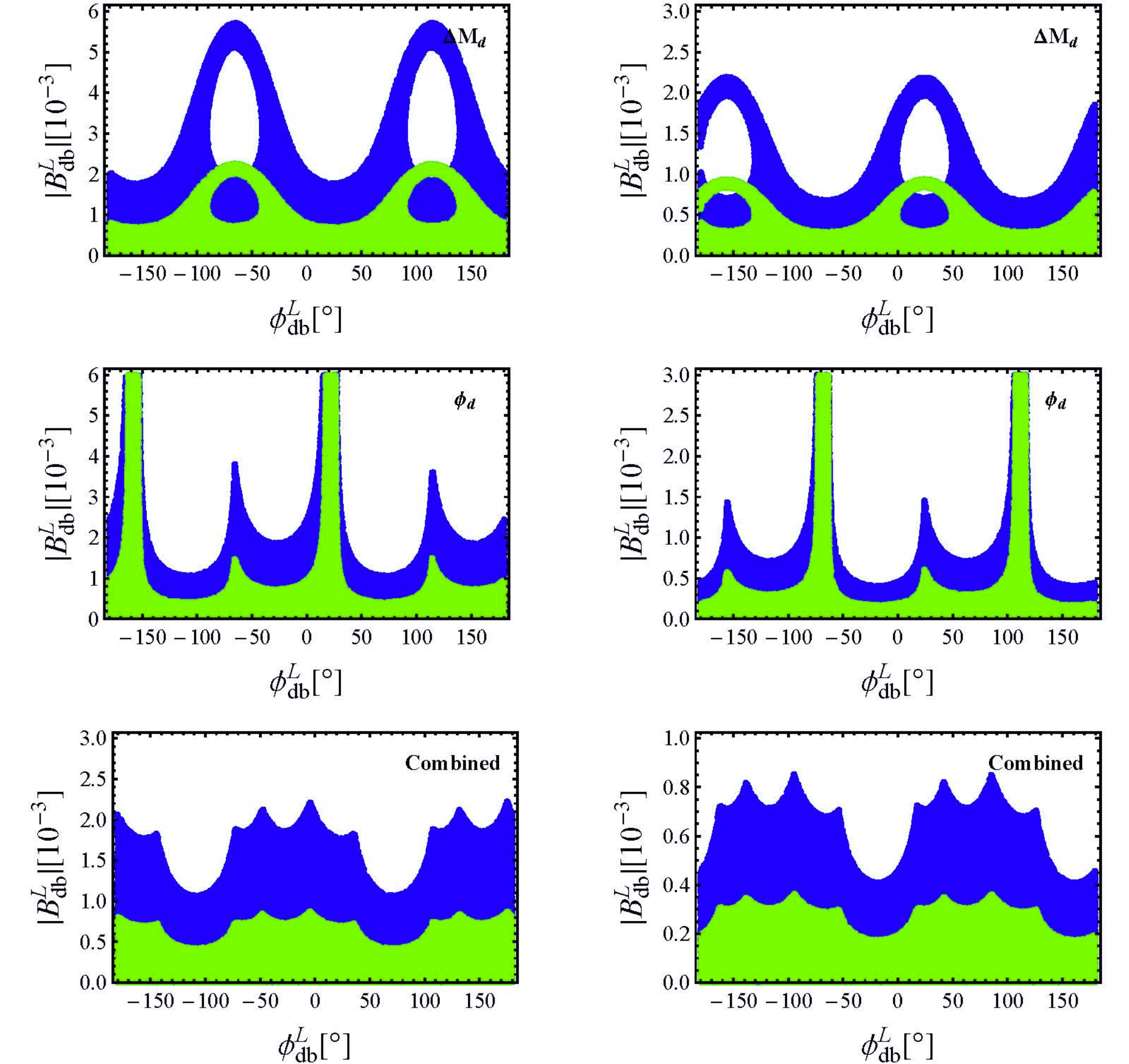}
\caption{\label{fig:Bd-mixing-final} \small The $\phi_{db}^{L}$-$|B_{db}^{L}|$ parameter space allowed by the $B_d^0-\bar B_d^0$ mixing observables $\Delta M_d$ and $\phi_d$. The left and the right panel correspond to the case with $B_{db}^{L}$ arbitrary and $B_{db}^{R}=0$, as well as the case with the simplification $B_{db}^{L}=B_{db}^{R}$, respectively. The green and the blue areas are obtained by choosing a $Z^\prime$ boson with $m_{Z^\prime}=400~{\rm GeV}$ and $m_{Z^\prime}=1~{\rm TeV}$, respectively.}
\end{figure}

Following the same procedure as in $B_s^0-\bar B_s^0$ mixing, our final results are shown in Fig.~\ref{fig:Bd-mixing-final}, where the left and the right panel correspond to the case with $B_{db}^{L}$ arbitrary and $B_{db}^{R}=0$~(the opposite case with $B_{db}^{L}=0$ and $B_{db}^{R}$ arbitrary is the same), as well as the case with the simplification $B_{db}^{L}=B_{db}^{R}$, respectively. The green and the blue areas are obtained by choosing a $Z^\prime$ boson with $m_{Z^\prime}=400~{\rm GeV}$ and $m_{Z^\prime}=1~{\rm TeV}$, respectively.

As can be seen from Fig.~\ref{fig:Bd-mixing-final}, the allowed parameter space in the $\phi_{db}^{L}$-$|B_{db}^{L}|$ plane is greatly reduced with the combined constraints from $\Delta M_d$ and $\phi_d$ imposed. It is also interesting to note that, similar to the hierarchy of the CKM matrix elements $|(V_{tb}V_{td}^{\ast})/(V_{tb}V_{ts}^{\ast})|\sim \mathcal{O}(10^{-1})$, the final allowed magnitudes $|B_{sb}^{L,R}|$ and $|B_{db}^{L,R}|$ satisfy roughly the same hierarchy, $|B_{db}^{L,R}|/|B_{sb}^{L,R}|\sim \mathcal{O}(10^{-1})$, within such a family non-universal $Z^\prime$ model. This may imply that the flavour-changing $Z^\prime$ couplings are also linked to the known structure of the SM Yukawa couplings~\cite{Chang:2009tx}.

\section{Conclusions}
\label{Sec:conclusion}

In this paper, motivated by the very recent measurements performed at the LHCb and the Tevatron of the mass difference $\Delta M_s$, the decay width difference $\Delta\Gamma_s$, the CP-violating phase $\phi_s$, as well as the like-sign dimuon charge asymmetry in semi-leptonic b-hadron decays, we have revisited the $B_s^0-\bar B_s^0$ mixing in a family non-universal $Z^{\prime}$ model, to check if the specific model could simultaneously explain the present measured values of these observables. Our main conclusions are summarized as follows:
\begin{enumerate}
\item[$\bullet$] In both data sets D1 and D2, the combination of measured mass difference $\Delta M_s$ and CP-violating phase $\phi_s$ could give a strong constraint on the flavour-changing $Z^\prime$ couplings $B_{sb}^{L,R}$, especially in the case where both the left- and right-handed currents are present at the same time.

\item[$\bullet$] In the first scenario where the $Z^\prime$ boson contributes only to the off-diagonal element $M_{12}^s$, we find that, once constraints from $\Delta M_s$, $\phi_s$ and $\Delta \Gamma_s$ are taken into account, the specific model could not reproduce the present measured value of the flavour-specific CP asymmetry $a_{fs}^s$ within its $1\sigma$ ranges.

\item[$\bullet$] Motivated by the failure of the first scenario, we have then considered the second scenario where the $Z^\prime$ boson contributes both to $M_{12}^s$ and to $\Gamma_{12}^s$. For simplicity, we have assumed that the NP contribution to $\Gamma_{12}^s$ comes only from tree-level $Z^\prime$-induced $b\to c\bar{c}s$ operators, with only the presence of left-handed flavour-changing coupling $B_{sb}^L$.

\item[$\bullet$] In the second scenario, due to the presence of flavour-diagonal couplings $B_{cc}^{L,R}$, we have considered the constraint from the indirect CP asymmetry in $\bar{B}_d\to J/\psi K_S$, in addition to the ones from $\Delta M_s$ and $\phi_s$. With these constraints imposed, we find that a large correction to $\Gamma_{12}^s$ is already excluded. While the predicted value of $\Delta \Gamma_s$ is roughly compatible with the experimental data, the measured $1\sigma$ experimental ranges of $a_{fs}^s$ still could not be reproduced in such a NP model.

\item[$\bullet$] For completeness, we have also presented the $Z^\prime$ effects on the mixing observables $\Delta M_d$ and $\phi_d$, assuming that the NP contribution comes only from the flavour-changing couplings $B_{db}^{L,R}$. It is interesting to note that, similar to the hierarchy of the CKM matrix elements, $|(V_{tb}V_{td}^{\ast})/(V_{tb}V_{ts}^{\ast})|\sim \mathcal{O}(10^{-1})$, the allowed magnitudes $|B_{sb}^{L,R}|$ and $|B_{db}^{L,R}|$ satisfy roughly the same hierarchy, $|B_{db}^{L,R}|/|B_{sb}^{L,R}|\sim \mathcal{O}(10^{-1})$. This may imply that the flavour-changing $Z^\prime$ couplings are also linked to the known structure of the SM Yukawa couplings~\cite{Chang:2009tx}.

\end{enumerate}

In conclusion, the specific $Z^\prime$ model we are considering could not simultaneously explain the measured $B_s^0-\bar B_s^0$ mixing observables, especially the like-sign dimuon charge asymmetry observed by the D0 collaboration. Future improved measurements from the LHCb and the proposed superB experiments, especially of the flavour-specific CP asymmetries, are expected to shed light on the issue.

\section*{Acknowledgements}

The work was supported in part by the National Natural Science Foundation of China~(NSFC) under contract Nos.~11005032, 11047125 and 10979008, and by the Specialized Research Fund for the Doctoral Program of Higher Education of China~(Grant No.~20104104120001). X.~Q. Li was also supported in part by MEC (Spain) under Grant FPA2007-60323 and by the Spanish Consolider Ingenio 2010 Programme CPAN (CSD2007-00042).

\begin{appendix}

\section*{Appendix: Theoretical input parameters}
\label{Sec:appendix}

In this appendix, we collect all the relevant input parameters for the $B_q^0-\bar B_q^0$ mixing observables both within the SM and in the family non-universal $Z^\prime$ model.

\subsubsection*{$\bullet$ The basic SM parameters}

First, we need some basic SM parameters, which are, if not stated otherwise, taken from the Particle Data Group~\cite{Nakamura:2010zzi}
\begin{align}
& \alpha_s(m_Z)=0.1184\pm 0.0007, \quad G_F=1.16637\times 10^{-5}~{\rm GeV}^{-2}, \quad \sin^2\theta_W=0.23146, \nonumber \\
& m_Z=91.1876~{\rm GeV}, \quad m_W=80.399~{\rm GeV}, \quad m_t=173.2\pm 0.9~{\rm GeV}~\cite{arXiv:1107.5255}, \nonumber \\
& m_{B_d}=5279.50~{\rm MeV}, \quad m_{B_s}=5366.3~{\rm MeV},
\end{align}
where $m_t$ is the top-quark pole mass, and we use two-loop running for $\alpha_s$ throughout this paper.

\subsubsection*{$\bullet$ The CKM matrix elements}

To discuss possible NP effects, we should treat the SM contribution as a theoretical background, and calibrate the CKM matrix elements exclusively on SM tree-level observables, which are largely insensitive to NP contributions. These include the moduli of CKM matrix elements from super-allowed $\beta$ decays, leptonic and semileptonic meson decays, as well as the CP-violating phase angle $\gamma$ from tree-dominated B-meson decays~\cite{Nakamura:2010zzi}. By fitting these constraints to the Wolfenstein parametrization~\cite{Wolfenstein:1983yz} of CKM matrix up to $\mathcal{O}(\lambda^4)$, it is found that~\cite{Dorsner:2011ai}
\begin{equation}
A=0.799\pm 0.026\,, \quad \lambda=0.22538\pm 0.00065\,, \quad \rho=0.124\pm 0.070\,, \quad \eta=0.407\pm 0.052\,,
\end{equation}
which are used through this paper.

\subsubsection*{$\bullet$ The heavy- and light-quark masses}

For the light-quark masses, we adopt the values determined by the Flavianet Lattice Averaging Group~(FLAG)~\cite{Colangelo:2010et}
\begin{equation}
\bar{m}_{s}(2~{\rm GeV})=94\pm 3~{\rm MeV}, \quad \bar{m}_{d}(2~{\rm GeV})=4.67\pm 0.20~{\rm MeV}.
\end{equation}
The scale-invariant $b$- and $c$-quark masses are taken as~\cite{Nakamura:2010zzi}
\begin{equation}
\bar{m}_b(\bar{m}_b)=4.19_{-0.06}^{+0.18}~{\rm GeV}, \quad \bar{m}_c(\bar{m}_c)=1.29_{-0.11}^{+0.05}~{\rm GeV}.
\end{equation}
To get the corresponding pole and running masses at different scales, we use the NLO $\overline {\rm MS}$-on-shell conversion and running formulae collected, for example, in Ref.~\cite{Chetyrkin:2000yt}.

For the mass parameter $m_b^{\rm pow}$, which appears in the $1/m_b$-suppressed matrix elements for the off-diagonal element $\Gamma_{12}^{s}$, we take $m_b^{\rm pow}=4.7\pm 0.1~{\rm GeV}$~\cite{Lenz:2006hd}.

\subsubsection*{$\bullet$ The non-perturbative parameters}

For the B-meson decay constants, we use the averaged lattice results performed by the CKMfitter group~\cite{Charles:2004jd}
\begin{equation}
f_{B_{s}}=231\pm 3\pm 15~{\rm MeV}, \quad f_{B_{s}}/f_{B_{d}}=1.235\pm 0.008\pm 0.033,
\end{equation}
where the first error is statistical and accountable systematic, while the second stands for systematic theoretical uncertainties~\cite{Charles:2004jd}.

Among the non-perturbative bag parameters $B_i^a$, the parameter $B_1^{VLL}$ is already known from the SM analyses, and we use~\cite{Lenz:2010gu}
\begin{align}
B_{1,B_s}^{VLL}(\mu_b)=0.841\pm 0.013\pm 0.020, \quad B_{1,B_s}^{VLL}/B_{1,B_d}^{VLL}=1.01\pm 0.01\pm 0.03,
\end{align}
renormalized at the scale $\mu_{b}=4.6~{\rm GeV}$. Since the parameter $B_1^{VRR}$ is the same as $B_1^{VLL}$, we can get the former from the latter. For the SM contribution, on the other hand, the RG invariant parameters $\hat{B}_{B_q}$ are adopted, and numerically we use~\cite{Charles:2004jd}
\begin{align}
\hat{B}_{B_s}=1.291\pm 0.025 \pm 0.035, \quad \hat{B}_{B_s}/\hat{B}_{B_d}=1.024\pm 0.013 \pm 0.015,
\end{align}
obtained by averaging over the lattice results.

To obtain the coefficient $P_1^{LR}(\mu_{Z^\prime})$, we need the bag parameters $B_1^{LR}$ and $B_2^{LR}$, the values of which are taken as~\cite{Becirevic:2001xt}
\begin{align}
& B_{1,B_s}^{LR}(\mu_b)=1.79\pm 0.04 \pm 0.18, \quad B_{1,B_s}^{LR}/B_{1,B_d}^{LR}=1.01\pm 0.03, \nonumber \\[0.2cm]
& B_{2,B_s}^{LR}(\mu_b)=1.14\pm 0.03 \pm 0.06, \quad B_{2,B_s}^{LR}/B_{2,B_d}^{LR}=1.01\pm 0.02.
\end{align}

To predict the off-diagonal element $\Gamma_{12}^s$, we also need the following bag parameters~\cite{Lenz:2010gu,Lenz:2006hd}
\begin{align}
& \tilde{B}_{S}(\mu_b)=0.91\pm 0.03\pm 0.12, \quad B_{R_{0,1}}(\mu_b)=B_{\tilde{R}_{1,2,3}}(\mu_b)=1.0\pm 0.5,\nonumber \\[0.2cm]
& B_{R_2}(\mu_b)=B_{\tilde{R}_{2}}(\mu_b), \quad B_{R_3}(\mu_b)=\frac{5}{7}B_{\tilde{R}_{3}}(\mu_b)+\frac{2}{7}B_{\tilde{R}_{2}}(\mu_b).
\end{align}

\subsubsection*{$\bullet$ The mass of $Z^\prime$ boson}

Since we are mainly concerned with the $Z^\prime$ couplings to fermions, we shall specify the $Z^\prime$-boson mass. At present, the best bound is provided by the CDF measurement~\cite{Acosta:2005ij}, which rules out a $Z^\prime$ boson with a mass below $399~{\rm GeV}$ at $95\%$ C.L.. As a guideline, we consider both a relatively light $Z^\prime$-boson with $m_{Z^\prime}=400~{\rm GeV}$, as well as a heavier one with $m_{Z^\prime}=1~{\rm TeV}$.

\end{appendix}

\end{document}